\documentclass[conference]{IEEEtran}
\usepackage{spconf,amsmath,graphicx}
\usepackage{epsf}

\usepackage{epsfig}

\usepackage{epstopdf}
\usepackage{cite}

\usepackage[dvips]{color}
\usepackage{epsf}
\usepackage{times}
\usepackage{epsfig}
\usepackage{graphicx}
\usepackage{url}
\usepackage{amsmath}
\usepackage{amssymb}
\usepackage{amsxtra}
\usepackage{algorithmic}
\usepackage{algorithm}
\usepackage{enumerate}

\newtheorem{theorem}{\bf Theorem}

\newtheorem{definition}{\bf Definition}

\DeclareMathOperator*{\sgn}{\,sgn}

\title{Integrating Energy Storage into the Smart Grid: \\
A Prospect Theoretic Approach\vspace{-0.15cm}}
\name {Yunpeng Wang$^1$, Walid Saad$^1$, Narayan B. Mandayam$^2$, H. Vincent Poor$^3$\vspace{-0.45cm}} 
\address{\small $^1$ Electrical and Computer Engineering Department, University of Miami, Coral Gables, FL, USA\\
\small $^2$ Department of Electrical and Computer Engineering, Rutgers University, North Brunswick, NJ, USA\\
\small $^3$ Electrical Engineering Department, Princeton University, Princeton, NJ, USA\\
\small Emails: \url{y.wang68@umiami.edu, walid@miami.edu, narayan@winlab.rutgers.edu, poor@princeton.edu}
\vspace{-0.65cm}}

\IEEEoverridecommandlockouts	

\begin{document}
\ninept
\maketitle
%%%%%%%%%%%%%%%%%%%%%%%%%%%%%%%%%%%%%%%%%%%%%%%%%%%%%%%%%%%%%%%%%%%%%%%%%%%%%%%%%%%%%%%%%%%%%%%%%%%%%%%%%%%%%%%%%%%%%%%%%%%%%%%%%%%%%%%%%%%%%%%%%%%%%%%%%%%%%

\begin{abstract}
In this paper, the interactions and energy exchange decisions of a number of geographically distributed storage units are studied under decision-making involving end-users. In particular, a noncooperative game is formulated between customer-owned storage units where each storage unit's owner can decide on whether to charge or discharge energy with a given probability so as to maximize a utility that reflects the tradeoff between the monetary transactions from charging/discharging and the penalty from power regulation. Unlike existing game-theoretic works which assume that players make their decisions rationally and objectively, we use the new framework of prospect theory (PT) to explicitly incorporate the users' subjective perceptions of their expected utilities. For the two-player game, we show the existence of a proper mixed Nash equilibrium for both the standard game-theoretic case and the case with PT considerations. Simulation results show that incorporating user behavior via PT reveals several important insights into load management as well as economics of energy storage usage. For instance, the results show that deviations from conventional game theory, as predicted by PT, can lead to undesirable grid loads and revenues thus requiring the power company to revisit its pricing schemes and the customers to reassess their energy storage usage choices.
%Simulation results show that incorporating user behavior via PT reveals several important insights on the prospective benefits and costs of integrating storage units, for both the power company and the customers.
\end{abstract}\vspace{-0cm}
\vspace{-0.2cm}

\section{Introduction}\vspace{-0cm}

Customer participation in energy management is seen as an integral feature of the smart grid~\cite{hossain2012smart}. In particular, the introduction of customer-owned storage units will provide the means for active user participation in managing energy transactions in the grid. For instance, these storage units provide the grid with the opportunity of storing energy at customer premises and they also allow customers to sell any surplus of energy available at their premises~\cite{hossain2012smart}. This represents a key feature for deploying smart grid applications such as demand response~\cite{mohsenian2010autonomous, saad2012game, logenthiran2012demand, macedo2013opportunities, atzeni2013noncooperative}. %Numerous studies have assessed the influence of deploying and maintaining storage units such as in [\textcolor{red}{ref}].  

%Storage units units have been suggested as technologies capable of exchanging energy within local area and such application faces many challenges. 
The integration of storage units into the smart grid, particularly at the customer side, requires overcoming many technical challenges ~\cite{5768106, garcia2008stochastic, pang2012bevs, 2011arXiv1104.3802A, weaver2009game, coogan2013energy, cui2013game}. The authors in ~\cite{5768106} addressed the problem of intermittent renewable energy generation by using energy storage to deal with dynamic loads and sources. In~\cite{garcia2008stochastic}, the authors studied the use of storage units as a means for complementing the stochastic generation of wind farms. In this work, the authors also investigate the impact of the presence of such storage units on the market price. 
%The work in \cite{pang2012bevs} investigated the use of plug-in hybrid electric vehicles as a dynamically dispersed energy storage for both demand side management and for outage management during faults. The authors in \cite{weaver2009game} presented a game theoretic approach to analyze and control the sources and loads in a small-scale DC system. In \cite{5356176}, local-scale energy consumption was minimized by coordinately charging, which could evaluate grid reinforcements and improve power quality. The work in \cite{2011arXiv1104.3802A} proposed a noncooperative game in which users could control their electricity consumption using two real-time pricing schemes. This work showed how the real-time cost of power generation impacts the customers consumption during peak hours. The authors in \cite{5540263} propose an approach to reduce the peak-to-average (PAR) ratio in smart meters using energy consumption schedule. The work in~\cite{dunn2011electrical} assess some battery technologies whose development offer low cost and is being applied in energy storage. The work in~\cite{strbac2008demand} reported that customers expect a reward instead of controlling penalty and such scheme was implemented by price-based signals. The work in \cite{erol2010using}, storage units were applied in a residential energy management by communication technologies to reduce the consumer bills. 
Other related game-theoretic solutions for smart grid pricing and energy management are discussed in \cite{2011arXiv1104.3802A, weaver2009game, coogan2013energy, cui2013game}.

Game theory has been a popular tool for smart grid design. However, most existing works assume that customers will abide by the rules of the game and act in a rational manner \cite{weaver2009game, 2011arXiv1104.3802A, mohsenian2010autonomous, saad2012game, coogan2013energy, cui2013game, 6193526, 6471883}. Indeed, none of these works incorporates the realistic behavior of the users which, in practice, can deviate from the conventional, rational norm set by game theory as observed in \cite{fiegenbaum1988attitudes, barberis2001prospect, levy1997prospect}. In this respect, prospect theory (PT), a Nobel-prize winning theory, provides the needed tools to explain how real-life user decisions can deviate from those predicted by conventional game theory \cite{kahneman1979prospect, tversky1981framing, tversky1992advances}. In particular, PT has shown that, in real life, users often act irrationally when faced with risk and uncertainty of outcome, as is the case in the smart grid, where the decisions of the customers are largely interdependent leading to risky outcomes.  These irrational decisions can stem not only from the users' behavior but also from computational errors occurring at the smart grid devices that are often resource-constrained. There are many studies that have applied PT to solve problems in the social sciences \cite{kahneman1979prospect, gao2010adaptive, harrison2009expected} as well as recent efforts to study the influence of end-users on wireless networks~\cite{6310922, Mandayam, clark2005tussle, wang2012studying, okuda2009design}. However, our work here is the first to break new ground in using PT to study end-user influence on the workings of the smart grid.

%However, few of the studies reporting these associations were prospective and none were designed to assess whether factors present early in the disease course predicted WD at a later stage. 
%However, few of the studies report users' perceptions and choices, and in reality the actual behavior of these users can deviate from the conventional, rational norm set by game theory. 
%For instance, recent studies \cite{kahneman1979prospect} show that these deviations can strongly affect the results predicted by game theory. 

The main contribution of this paper is to propose a new framework for energy management in the smart grid using the tools of prospect theory. In particular, we formulate a noncooperative game between the customer-owned storage units, in which the decision of each customer explicitly incorporates its \emph{subjective} perception on the actions taken by other customers. In this game, each customer can decide whether to charge its storage unit or sell the available surplus to the grid, while optimizing a utility that captures the associated costs and benefits, under a subjective observation of the other customers' actions. Compared to related works on smart grid markets and demand response \cite{mohsenian2010autonomous, saad2012game, logenthiran2012demand, macedo2013opportunities, 5768106, garcia2008stochastic, pang2012bevs, 2011arXiv1104.3802A, weaver2009game, coogan2013energy, cui2013game, 6193526, 6471883}, our paper has several new contributions: \emph{1)} in contrast to the conventional expected utility theory (EUT), we develop a novel PT-based framework that allows proper modeling of realistic user behavior during energy management; \emph{2)} we design a novel game-theoretic model that allows incorporation of both economic (pricing) and power factors (grid regulation); and \emph{3)} we show the existence of a mixed Nash equilibrium for the proposed game under PT considerations. %In particular, we compare the revenue of power company through the utility of players, in which each one could determine underlying action with objective probability or subjective estimation. %We are particularly interested in overcoming two key challenges: (a) introducing a new approach using subjective observation to decide on whether charge or discharge energy while while taking into account the effect of power regulation from the sider of generation, and (b) developing and analyzing a existence of a mixed NE under ET and PT that involves a weight representing human feelings. 
Extensive simulation results show that deviations from the rational, EUT behavior can lead to unexpected, and possibly undesirable performance, in terms of power company revenues and average grid load.

The remainder of the paper is organized as follows: Section~\ref{sec:sysmodel} presents the studied system model and formulates the problem as a game with PT considerations. In Section~\ref{sec:solution}, we analyze the equilibrium for the two-player case. Simulation results are presented in Section~\ref{sec:sim}, while conclusions are drawn in Section~\ref{sec:conc}.\vspace{-0.1cm}

\section{System Model and PT Game Formulation}\label{sec:sysmodel}\vspace{-0.1cm}
Consider a smart grid in which $N$ customers are present. Let $\mathcal{N}$ be the set of all $N$ customers. Under normal operating conditions, we assume that each customer $i \in \mathcal{N}$ constitutes a constant load $D_i$ on the grid. Among all $N$ customers, a subset $\mathcal{K} \subseteq \mathcal{N}$ of $K$ customers is assumed to be ``active''. Here, an active customer refers to a user equipped with a smart home and able to actively participate in the energy management of the smart grid, as allowed by the power company. Every customer $k \in \mathcal{K}$ owns a storage unit that initially stores an amount of energy $S_k\! <\! D_k$. At a given period of time, we assume that the participation of each customer $k \in \mathcal{K}$ is restricted to one of two actions: a) charge the needed amount $D_k$ (act as load) or b) discharge/sell the surplus $S_k$ to the other customers (act as source).
%For ease of analysis

Naturally, any given action by a customer $k \in \mathcal{K}$ will affect both the power system (needed generation, losses, etc.) and the market economics (prices). We assume that the power company allows the customers to charge or discharge, but it requires that the total generation power remains within a nominal, desirable value to maintain the power system's stability \cite{glover2008power}. %On the one hand, the revenue of power company consists fixed costs and variable costs. Fixed costs relate to the monies that have to be spent even no power generates, including capital costs, taxes and operations costs. Variable costs are the necessary cost for running the plant to generate power.
%Thus, for each participating user, they would buy or sell the energy based on the total generation power and their exchange power would also impact the system, in which traditionally releases energy on peak-hour or seasonal requirements. For such influence, i.e., power regulation, each user $k \in \mathcal{K}$ exchanges some energy $E_k$ in a short time:
%\begin{equation}
%G-\sum_k E_k >0
%\end{equation}
%with $G$ being the maximum total power that the power company has to provide in market. %A rational person might consider three perspectives when a real-time energy/power information appears on a smart meter: the benefit, the price and the effect of latent loss. 
%For all $N$ users, they would pay for the energy they needed and the loss over transmission line; 
In this studied scenario, all $K$ participating users use storage units to charge and discharge so as to optimize their overall monetary benefits. The decisions of the customers are, however, largely coupled, which leads to a game-theoretic setting as discussed next.\vspace{-0.1cm}

\subsection{Noncooperative Game Model}\vspace{-0.05cm}
We analyze the interactions between the active customers using noncooperative game theory~\cite{GT00}. %In this paper, we assume that the strategy choices of the customers are largely \emph{interdependent}, but each player could estimate others' possible actions based on a personal perception. Consequently, we 
As the strategy choices of the customers are largely \emph{interdependent}, we can formulate a strategic noncooperative game $\Xi=(\mathcal{K},\{\mathcal{A}_k\}_{k\in\mathcal{K}},\{u_k\}_{k\in \mathcal{K}})$, that is characterized by three main
elements: \emph{a)} the players are the active customers in the set $\mathcal{K}$, \emph{b)} the action $a_k \in \mathcal{A}_k := \{D_k,S_k\}$ of each player is to either charge/buy a total amount of energy $D_k$ ($a_k=D_k$) or discharge/sell the available surplus $S_k$ $(a_k=S_k)$, and \emph{c)} the utility function $u_k$ of each player $k$ which captures the benefit-cost tradeoffs associated with the different choices.  Each customer $k$ is assumed to have enough storage capacity to handle an amount $D_k + S_k$. Here, we note that, although the customers may have other demands, our model is solely focused on the discharge/charge actions and their impact on the grid and customers. The utility function achieved by a player $k \in \mathcal{K}$ that chooses an action $a_k$ is given by \vspace{-0.1cm}
\begin{equation}\footnotesize \label{eq:util}\vspace{-0.1cm}
\begin{split}
u_k(a_k, \boldsymbol{a}_{-k})=& - \alpha(a_k,\boldsymbol{a}_{-k}) \Bigl( D_k + L_k(a_k,\boldsymbol{a}_{-k}) \Bigr) \\& + \gamma(a_k,\boldsymbol{a}_{-k}) S_k -\beta \Bigl(  G(a_k, \boldsymbol{a}_{-k})-\hat G \Bigr)^2,
\end{split}
\end{equation}
where $\boldsymbol{a}_{-k}=[a_1, a_2, \dots, a_{k-1}, a_{k+1}, \dots, a_K]$ is the vector of action choices of all players other than $k$,  $L_k(a_k,\boldsymbol{a}_{-k})$ are the total losses over the distribution/transmission lines which depend on the total demand and are computed using conventional optimal power flow algorithms \cite{glover2008power}, $G(a_k, \boldsymbol{a}_{-k})$ denotes the total generation by \emph{the power company} (not the customers) under current action choices, and $\beta$ is a regulation penalty factor, that allows the power company to maintain a regulated power supply, i.e. $\hat G$. Maintaining such a regulation is important for many operational aspects of the grid, such as the conversion between AC and DC. We note that, in our game, the actions are positive and we have positively/negatively defined charging/discharging unit payments $\alpha$ and $\gamma$ in (\ref{eq:util}). Here, we define the charging price and the discharging price, respectively, set by the power company and participating users as follows:\vspace{-0.1cm}
\begin{equation}\footnotesize \label{eq:utilcharge}\vspace{-0.1cm}
\alpha(a_k,\boldsymbol{a}_{-k})  = \begin{cases}
c(a_k,\boldsymbol{a}_{-k}) &\textrm{ if } a_k = D_k,\\
0&\textrm{ otherwise,}
\end{cases}
\end{equation}
with $c(a_k,\boldsymbol{a}_{-k})$ being the unit price in the energy market which follows the pricing strategy of the power company. Moreover,\vspace{-0.1cm}
\begin{equation}\footnotesize \label{eq:utildischarge}\vspace{-0.1cm}
 \gamma(a_k,\boldsymbol{a}_{-k})  = \begin{cases}
b_k &\textrm{ if } a_k = S_k,\\
0&\textrm{ otherwise,}
\end{cases}
\end{equation}
with $b_k$ being the unit price at which a certain customer $k$ would sell its surplus $S_k$. We assume that each customer can set its own price, but the power company will impose a pricing restriction $B$, such that $b_k < B,\ \forall k \in \mathcal{K}$. 

The utility function in (\ref{eq:util}) captures both the economic benefits of customer participation as well as the impact on the power system (via the regulation term). Here, while the power company allows the $K$ active customers to actively decide on whether to buy or sell energy, it mandates that the generated power in the considered geographical area remains within desired, stable operating conditions. Also, we note that both demand and line loss determine the total generation level and, excessive charging or discharging might damage the generator due to a frequency variation thus requiring regulation~\cite{begovic1993frequency}. Without loss of generality, we assume that the normal, stable operating conditions correspond to the case in which all $N$ customers act as loads and we let $\hat{G} = \text{\footnotesize $\sum$}_{i\in \mathcal{N}} (D_i + L_i)$ denote the total generated power required for this distribution area during normal operation. $L_i$ represents the losses incurred over the distribution/transmission lines for delivering $D_i$ to customer $i$ which depend on the total demand and are computed using power flow algorithms. Therefore, for the case in which $a_k=D_k, \forall k \in \mathcal{K}$, we have $G(\boldsymbol{a})=\hat{G}$ (with $\boldsymbol{a}$ being the vector of all strategies). Consequently, any actions taken by a certain customer that shifts the generated power from its nominal value $\hat{G}$ will require the power company to regulate the generation. The need for this regulation indirectly yields a cost penalty on the active participants as captured in (\ref{eq:util}).\vspace{-0.1cm}

\subsection{Expected Utility Theory}\vspace{-0.05cm}
In a smart grid, owing to uncertainty in power generation as well as the fact that the customers can make certain decisions (such as whether to allow the use of their storage device or not) with different frequency over time, it is reasonable to assume that customers make probabilistic choices. Therefore, we are interested in studying the game under \emph{mixed strategies}~\cite{GT00}. %using the conventional notions of expected utility theory ~\cite{GT00}. %Before we further explore a noncooperative game, we firstly study the mixed strategies via Expected utility theory [\textcolor{red}{ref}]. For a noncooperative game, participating users are informed the historic information regarding to others actions but they cannot communicate with each other. Without such players, or emerging charging/discharging storage unit, a increasing requirement enforce the growth of generation on the side of power plant so as to maintain the stability of power system. Active participants can obtain an individual benefit while also assisting the power company in maintaining stability by selling/discharging energy to the grid. However, the expected selling price $b_k$ and underlying discharging amount are individually different, and the discharging amount comes from a charging amount in which each player have to pay for the energy if they receive energy from power company. In other words, the payment not only relate to the amount of energy but the real-time electricity price, in which the power company announced by observing the total requirements. For player $k$, we denote $p_k \in \mathcal{P}$ as the probability of charging and $(1-p_k)$ as the probability of discharging. 
As customers are often uncertain when presented with different choices in practice, a mixed-strategy solution can better capture their realistic behavior. Let $\boldsymbol{p}= [p_1,\ldots,p_k]$ be the vector of all mixed strategies, where, for every customer $k\in \mathcal{K}$, $p_k(a_k)$ is the probability distribution over the pure strategies $a_k \in \mathcal{A}_k$. %Hereinafter, we let $p_k(D_k)$ be the probability that any customer $k \in \mathcal{K}$ chooses to charge, i.e., $a_k=D_k$, and $p_k(S_k)$ be the probability of discharging, i.e. $a_k=S_k$. 

Under the conventional EUT model, the utility of each user is simply the expected value over its mixed strategies. Thus, the EUT utility of a player $k$ is given by \vspace{-0.2cm}
\begin{equation}\footnotesize \label{eq:multiplayerET}\vspace{-0.2cm}
\begin{split}
&U_k^{\text{EUT}}( \boldsymbol{p})=\sum_{\boldsymbol{a} \in \mathcal{A}}\bigg(\prod_{l=1}^K p_l(a_l)\bigg) u_k(a_k, \boldsymbol{a}_{-k}),
\end{split}
\end{equation}
where $\boldsymbol{a}$ is the vector of all players' strategies and $\mathcal{A}=\mathcal{A}_1 \times \mathcal{A}_2 \times \dots \times \mathcal{A}_K$.\vspace{-0.1cm}% and $p_l(\cdot)$ is the probability distribution over the actions $a_l$, used by player $l$.\vspace{-0.1cm}

\subsection{Prospect Theory}\label{sec:pt}\vspace{-0.1cm}

As previously mentioned, EUT evaluates an objective expected utility in which users are assumed to act rationally and objectively. However, it has been observed that, in real life, users' behavior deviates considerably from the rational path predicted by EUT. %PT studies \cite{kahneman1979prospect} have shown that users have subjective evaluations of gains and losses when making decisions under risk. 
For the proposed game, a customer $k$ has to decide on its action, in the face of uncertainty induced by the mixed strategies of its opponents, which impact directly its utility as in (\ref{eq:multiplayerET}). In order to capture such behavioral factors in the proposed energy trading game, we turn to the framework of prospect theory \cite{kahneman1979prospect}. 

%ET formulates explicit numerical probabilities, and PT responses this explicit probabilities through adopting their stated values. A classic PT mainly defines a wight $w$ to evaluate the probability that each player observes and then, underlying average utility would be changed due to the effect of such weight. This effect reveals a subjective observation $w(p_m)$ rather than the objective probability $p_m$ so as to weigh a human natural attitudes of possible outcomes. Thus, decision weight is not a probability but a value to infer from choices between prospects. This decision weight accompanies with an event in which the objective probability could be influenced by other factors. For example, a certain event holds $w(1)=1$ and an impossible event holds $w(0)=0$, since the decision weight does not change the ``probability" scales of truly or impossible events. Excluding these two cases, most people could overly weigh low probability outcomes and underly weigh high probability outcomes due to a psychological result. Hence, in this paper, we use a weight in [\textcolor{red}{ref}] and capture the over and under weight of probabilistic outcomes as:
One important PT notion is the so-called \emph{weighting effect}. In particular, in PT~\cite{prelec1998probability} it is observed that in real-life decision-making, people tend to subjectively weight uncertain outcomes. In the proposed game, this weighting effect allows capture of each user's subjective evaluation on the mixed strategy of its opponents.  Thus, under PT, instead of objectively observing the mixed strategy vector $\boldsymbol{p_{-k}}$ chosen by the other players, each user perceives a weighted version of it, $w_k(\boldsymbol{p_{-k}})$. Here, $w_k(\cdot)$ is a nonlinear transformation that maps an objective probability to a subjective one. PT studies have shown that most people could often overweight low probability outcomes and overweight high probability outcomes \cite{kahneman1979prospect}. Hereinafter, we assume that all players utilize a similar weighting approach, such that $w_k(\cdot)=w(\cdot),\ \forall k \in \mathcal{K}$. While many weighting functions exist in the PT literature, we choose the popular Prelec function (for a given probability $\sigma$)~\cite{prelec1998probability}:
\vspace{-0.15cm}
\begin{equation}\footnotesize \label{eq:weight}\vspace{-0.15cm}
w(\sigma)=\exp(-(-\ln \sigma)^\alpha),\ 0<\alpha \le 1,
\end{equation}
where $\alpha$ is a parameter used to characterize the distortion between subjective and objective probability. Note that when $\alpha=1$, (5) is reduced to the conventional EUT probability. %Fig.~\ref{fig:probaweight} illustrates the probability weighting effect. In this figure, we can see that, the objective probability and subjective estimation insect at $p=1/e$ and the curve approximates to $w=1/e$ as $\alpha$ increases. 

%\begin{figure}[!t]
%  \begin{center}
 %  \vspace{-0.2cm}
  %  \includegraphics[width=7.5cm]{probaweight.eps}
 %  \vspace{-0.3cm}
%    \caption{\label{fig:probaweight} the effect of weight as the %objective probability varies.}
 % \end{center}\vspace{-0.9cm}
%\end{figure}

%Comparing to (\ref{eq:multiplayerET}), we could obtain a prospect utility using personal subjective evaluation/observation:
Under PT, the expected utility achieved by a player $k$, given the weighting effect, is \vspace{-0.3cm}
\begin{equation}\footnotesize \label{eq:multiplayerPT}\vspace{-0.2cm}
\begin{split}
&U_k^{\text{PT}}( \boldsymbol{p}) = \sum_{\boldsymbol{a} \in \mathcal{A}}\!\! \bigg(p_k(a_k)\!\!\!\!\!\! \prod_{l \in \mathcal{K} \setminus \{k\}}^K\!\!\!\!\!\! w(p_l(a_l))\bigg) u_k(a_k, \boldsymbol{a}_{-k}).
\end{split}
\end{equation}
%\begin{equation}\label{eq:multiplayerPT}
%\begin{split}
%&U_k^{\text{PT}}(p_1,p_2, \dots, p_K)=\boldsymbol{P}^{\text{PT}}\cdot \boldsymbol{u}\\
%&\boldsymbol{P}^{\text{PT}}=[w_k(p_1)\ w_k(1-p_1)]\otimes[w_k(p_2)\  w_k(1-p_2)]\otimes \cdots \\
%&\qquad \quad \otimes [p_k\  (1-p_k)] \otimes \dots \otimes [w_k(p_K)\  w_k(1-p_K)]\\
%&\boldsymbol{u}=[D_1\ S_1]^T\otimes[D_2\  S_2]^T\otimes  \dots \otimes [D_K\  S_K]^T
%\end{split}
%\end{equation}

Here, we assume that a player uses a subjective evaluation only on the other players' strategy probabilities. Thus, customer $k$'s subjective evaluation of its own probability is equal to its objective probability. %Next, we will study the solution of the game using the notion of the mixed-strategy Nash for both EUT and PT. 
Given the set of probability distributions $\mathcal{P}_k$ over its set of strategies $\mathcal{A}_k$, the solution of the game can be found via the notion of a mixed-strategy Nash equilibrium:
\begin{definition}
A mixed strategy profile $\boldsymbol{p}^* \in \mathcal{P}=\prod_{k=1}^K P_k $ is a mixed strategy Nash equilibrium if, for each player $k \in \{1,2,\dots, K\}$, we have (for either PT or EUT)\vspace{-0.2cm}
\begin{equation}\footnotesize
U_k(p_k^*,\boldsymbol{p}_{-k}^*) \ge U_k(p_k,\boldsymbol{p}_{-k}^*), \  \forall p_k \in  \mathcal{P}_k.
\end{equation}
\end{definition}\vspace{-0cm}

\section{Solution: The Two-Player Case}\label{sec:solution}\vspace{-0.05cm}
To gain greater insight into the solution of the proposed game, we analyze a case study for the scenario in which only $K=2$ customers are active. In particular, we are interested in analyzing the \emph{proper mixed Nash equilibrium} of the game. A proper mixed-strategy Nash equilibrium is the solution in which each player chooses a certain action $a_k$ with probability $0<p_k<1$. %Moreover, we assume that each player uses the PT weights only when evaluating the probability of the opponent. The two-player case can be represented by a matrix game, as shown in Table~\ref{tab:2player}.
While the existence of a mixed-strategy Nash equilibrium is well-known for conventional EUT games~\cite{GT00}, it is of interest to study whether the PT game admits such an equilibrium. Moreover, for both the EUT and PT games, we are interested in guaranteeing a \emph{proper} mixed strategy Nash equilibrium, in which the users will indeed mix between their strategies. With this in mind, we can state the following result:

\begin{theorem}\label{th:existence}
For the proposed two-player smart grid game $\Xi=(\mathcal{K},\{\mathcal{A}_k\}_{k\in\mathcal{K}},\{u_k\}_{k\in \mathcal{K}})$, there exists a unique, proper mixed Nash equilibrium for both the EUT and PT games if $-c(D_k,D_{-k}) D_k+\beta (D_k+S_k)^2<b_kS_k<-c(D_k,S_{-k}) D_k+\beta (D_k+S_k)^2+2\beta \prod_{l=1}^{2} (D_l+S_l)$, where $k=\{1,2\}$.
%\begin{equation}\label{eq:benefitcondition1}
%-c(D_1,D_2) D_1+\beta (D_1+S_1)^2<b_1S_1<-c(D_1,S_2) %D_1+\beta (D_1+S_1)^2+2\beta (D_1+S_1)(D_2+S_2)
%\end{equation}
%and 
%\begin{equation}\label{eq:benefitcondition2}
%-c(D_1,D_2)D_2+\beta (D_2+S_2)^2<b_2S_2<-c(S_1,D_2) D_2+\beta (D_2+S_2)^2+2\beta (D_1+S_1)(D_2+S_2).
%\end{equation}
\end{theorem}\vspace{+0.06cm}

\begin{proof}
In the proposed model, there always exists at least one mixed NE under EUT as guaranteed by Nash's result \cite{GT00}. Thus, our proof mainly focus on finding a condition to guarantee \emph{1)} there exists a \emph{proper} mixed NE under EUT and PT, and \emph{2)} such a proper mixed NE is unique. By using the indifference principle under EUT, a proper mixed-strategy Nash equilibrium, $(p_1^*, p_2^*)$, exists when the average charging utility is equal to the average discharging utility. For example, computing customer $1$'s average utility by $p_2^*$, we have $p_2^*u_1(D_1,D_2)+(1-p_2^*)u_1(D_1,S_2)=p_2^*u_1(S_1,D_2)+(1-p_2^*)u_1(S_1,S_2)$; that is, \vspace{-0.1cm}
\begin{equation}\footnotesize \vspace{-0.1cm}
p_2^*=\frac{u_1(S_1,S_2)-u_1(D_1,S_2)}{u_1(D_1,D_2)-u_1(S_1,D_2)+u_1(S_1,S_2)-u_1(D_1,S_2)}.
\end{equation}
A sufficient condition to have a proper mixed strategy Nash equilibrium, such that $0 < p_2^* < 1$, is to have
\vspace{-0.1cm}
\begin{equation}\label{eq:p2conditionET}\footnotesize \vspace{-0.1cm}
\sgn \biggl(u_1(S_1,S_2)-u_1(D_1,S_2)\biggr)\!=\!\sgn \biggl(u_1(D_1,D_2)-u_1(S_1,D_2)\biggr),
\end{equation}
where $\sgn(\cdot)$ denotes the algebraic sign of its argument and\vspace{-0.1cm}
\begin{equation}\! \! \! \! \!  \footnotesize \vspace{-0.1cm}
\begin{cases}
u_1(D_1,D_2)\!\!\!\!\!\!&=-c_{11}(D_1+L_1(D_1,D_2)), \\
u_1(D_1,S_2)\!\!\!\!\!\!&=-c_{12} (D_1+L_1(D_1,S_2))-\beta (G(D_1,S_2)-\hat G)^2,\!\!\!\!\!\!\\
u_1(S_1,D_2)\!\!\!\!\!\!&=b_1S_1-\beta (G(S_1,D_2)-\hat G)^2, \\
u_1(S_1,S_2)\!\!\!\!\!\!&=b_1S_1-\beta (G(S_1,S_2)-\hat G)^2.
\end{cases}
\end{equation}

On the other hand, we assume that player $i$'s subjective evaluation of its own probability is equal to its objective probability, such that $w_1(p_1)=p_1$ and $w_2(p_2)=p_2$. Then, using the indifference principle under PT, player $1$'s average utility of charging $w_1(p_2^*)u_1(D_1,D_2)+w_1(1-p_2^*)u_1(D_1,S_2)$ is equal to its average discharging utility $w_1(p_2^*)u_1(S_1,D_2)+w_1(1-p_2^*)u_1(S_1,S_2)$; that is, \vspace{-0.1cm}
\begin{equation}\label{eq:p2conditionPT}\footnotesize \vspace{-0.1cm}
\frac{w_1(p_2^*)}{w_1(1-p_2^*)}=\frac{u_1(S_1,S_2)-u_1(D_1,S_2)}{u_1(D_1,D_2)-u_1(S_1,D_2)}>0,
\end{equation}
which is analogous to the condition (\ref{eq:p2conditionET}) under EUT. Computing player $2$'s average utility by $p_1^*$, we also have the condition\vspace{-0.1cm}
\begin{equation*}\footnotesize \vspace{-0.1cm}
\sgn \biggl(u_2(S_1,S_2)-u_2(S_1,D_2)\biggr)\!=\!\sgn \biggl(u_2(D_1,D_2)-u_2(D_1,S_2)\biggr).
\end{equation*}

To solve (\ref{eq:p2conditionET}), we need to simplify\vspace{-0.1cm}
\begin{equation}\footnotesize \label{eq:onlyless1}\vspace{-0.1cm}
\begin{split}
&u_1(D_1,D_2)-u_1(S_1,D_2)\\
=&-c_{11}(D_1+L_1(D_1,D_2))-b_1S_1+\beta (G(S_1,D_2)-\hat G)^2,\\
=&-c_{11}(D_1+L_1(D_1,D_2))-b_1S_1+\beta \{[D_2-S_1+D_{\text{others}}\\
&+L(S_1,D_2)]-[D_1+D_2+D_{\text{others}}+L(D_1,D_2)]\}^2,\\
=&-c_{11}D_1-b_1S_1+\beta (D_1+S_1)^2,
\end{split}
\end{equation}
where $D_{\text{others}}$ represents the total constant demand of non-participating users. Here, we have assumed that the losses $L(\cdot)$ are negligible with respect to the demand, which is a reasonable assumption when dealing with two players only, i.e., $L_k << D_k,\ k=1,2$. Similarly, \vspace{-0.1cm}
\begin{equation}\footnotesize \label{eq:onlyless2}\vspace{-0.1cm}
\begin{split}
&u_1(S_1,S_2)-u_1(D_1,S_2)\\
=&b_1S_1-\beta (D_1+S_1+D_2+S_2)^2+c_{12} D_1+\beta (D_2+S_2)^2,\\
=&c_{12} D_1+b_1S_1-\beta (D_1+S_1)^2-2\beta (D_1+S_1)(D_2+S_2).
\end{split}
\end{equation}
If (\ref{eq:onlyless1}) is greater than 0, (\ref{eq:onlyless2}) cannot be greater than 0 due to the fact that, in practice, as the locational marginal pricing (LMP) \cite{powerbookLMP} increases with the generated power, the price at a lower generation level cannot exceed that charged at a higher level; thus, mathematically, $c_{12} \le c_{11}$. Thus, both sides of (\ref{eq:p2conditionET}) have to be negative and then, we obtain the range of $b_kS_k$ in Theorem~\ref{th:existence}.
\end{proof}
%\remark The utility function is not a pure monetary due to the power regulation part. Thus, the ``pure" revenue of a player is not equal to its utility.
%\remark We have proved the existence and uniqueness of a mixed NE based on a [\textcolor{red}{fixed reference}].

Under given loads and surpluses, Theorem~\ref{th:existence} provides a relationship between the unit selling price $b_k$ of each player, the LMP price $c(a_1,a_2)$, and the penalty factor for regulation $\beta$, such that we could obtain a proper mixed strategy equilibrium. From the utility functions in (\ref{eq:multiplayerET}) and (\ref{eq:multiplayerPT}), we can mathematically see the difference between EUT and PT. Here, given the players' mixed strategies, we define the company's expected revenues under the equilibrium probabilities, $(p_1^*,p_2^*)$ for EUT and $(p_1^{*,\textrm{PT}},p_2^{*,\textrm{PT}})$ for PT. The power company generates a revenue depending on the energy sold to the two customers, although the customers' probability of charging or discharging can be different between EUT and PT. Thus, the power company revenues obtained from customers $1$ and $2$ are as follows:\vspace{-0.1cm}
\begin{equation}\footnotesize \label{eq:companymoney}\vspace{-0.1cm}
\begin{split}
\!\!\!\!\!\!R_{EUT}\!\!=&p_1^*p_2^*c_{11}(D_1+D_2+L_{1,2})+p_1^*(1-p_2^*)c_{12}(D_1+L_1)\\
&+(1-p_1^*)p_2^*c_{21}(D_2+L_2),\\
\!\!\!\!\!\!R_{PT}\!\!=&p_1^{*,\textrm{PT}}p_2^{*,\textrm{PT}}c_{11}(D_1+D_2+L_{1,2})+p_1^{*,\textrm{PT}}(1-p_2^{*,\textrm{PT}})c_{12}\\
&(D_1+L_1)+(1-p_1^{*,\textrm{PT}})p_2^{*,\textrm{PT}}c_{21}(D_2+L_2),
\end{split}
\end{equation}
where $L(\cdot)$ is the loss in power flow (\ref{eq:util}). $R_{EUT}$ is the expected revenue obtained by the power company. And $R_{PT}$ is the PT revenue obtained by the power company, in which player $1$ and player $2$ use their subjective perspectives.\vspace{-0.1cm}
%; in other words, the company would get revenue by naturally considering individual, subjective probability. Here, we remarked that $w(1-p_1^{PT})$ in (\ref{eq:2playerPT}) and $(1-p_1^{PT})$ in (\ref{eq:companymoney}) are different: the weight is used for computing individual utility consisting of charging, discharging and payment of power regulation; the subjective probability is used to compute an average revenue via containing the average transmitting energy with underlying price.

\section{Simulation Results and Analysis}\label{sec:sim}\vspace{-0.1cm}

For simulating the proposed system, we consider a geographical region in which two active customers equipped with storage units exist. %two-player (i.e. smart buildings and PHEVs parking plots) could have stored energy that they wish to sell to the existing demand (i.e. constant load) in smart grid.
We choose typical values for the demand and surplus: $D_1=20\textrm{ kWh}, D_2=15\textrm{ kWh}, S_1=10\textrm{ kWh}, S_2=5\textrm{ kWh}, \alpha=0.25, \beta=0.0018$. The constant load is set as 200\textrm{ kWh}, and power line parameters are set from a typical 4-bus system \cite{grainger2003power}. The following examples assume that the generation power (kW) is numerically equal to the energy (kWh) in a one-hour time unit. %In the following numerical examples, we show how the behavior of the customers under PT can deviate from EUT and, thus, yield desirable or undesirable conditions in the overall energy market. 
For pricing, we assume that $c(a_1,a_2)$ follows a conventional LMP scheme, such as the following:\vspace{-0.2cm}
\begin{equation*}\footnotesize  \label{eq:price}
c\!=\! \begin{cases}
$\$0.05$/\textrm{kWh} &\! \textrm{power company generation is $\le$ } $200$~\textrm{kWh}  ,\\
$\$0.10$/\textrm{kWh}  &\! \textrm{power company generation between }  $200$-$250$~\textrm{kWh},\\
$\$0.15$/\textrm{kWh} &\! \textrm{power company generation between } $250$-$300$~\textrm{kWh}  ,\\
$\$0.20$/\textrm{kWh}  &\! \textrm{power company generation is $>$ } $300$~\textrm{kWh} .\\
\end{cases}
\end{equation*}

\begin{figure}[!t]
  \begin{center}
   \vspace{-0.3cm}
    \includegraphics[width=7.5cm]{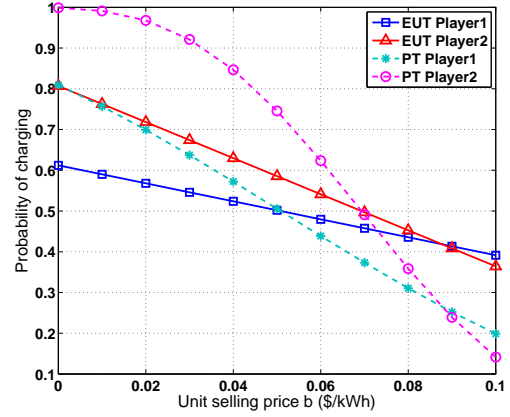}
   \vspace{-0.4cm}
    \caption{\label{fig1} Probability of charging under EUT and PT as $b$ varies.}
  \end{center}\vspace{-0.9cm}
\end{figure}

\vspace{-0.2cm}In Fig.~\ref{fig1}, we depict the impact of the unit selling price on the behavior of the customers. Without loss of generality, we assume that both customers use the same price $b_1=b_2=b$ and we vary the price within the range in which the equilibrium exist as per Theorem~\ref{th:existence}. Fig.~\ref{fig1} shows how the probability of charging for both players varies as $b$ increases, for both the EUT and PT cases. Clearly, as the selling price increases, both players would have more incentive to discharge than to charge, as the benefit would start outweighing the regulation penalty. More interestingly, Fig.~\ref{fig1} shows that, for both customers, the PT behavior significantly differs from the EUT behavior. For example, for customer $2$, below a selling price of $b=\$ 0.07$ per kWh, the probability of charging at the equilibrium for PT is much higher than EUT. This implies that for low gains, each customer follows a more \emph{conservative, risk-averse} strategy under PT and is less interested in reaping the benefits of selling energy than in the EUT case. However, as the selling price crosses the threshold, the probability of charging for customer $2$ under PT becomes much smaller than under EUT. This implies that once the selling benefits are significant (and the risks decrease), customer $2$ starts selling more aggressively under PT than under EUT. A similar behavior can be observed for customer $1$, although the benefit threshold of customer $1$ is smaller ($b=\$ 0.05$), since customer $1$ has more energy to sell/buy.
%(which is comparable to the minimum charging price set by the power company), 

\begin{figure}[!t]
 \begin{center}
  \vspace{-0.3cm}
   \includegraphics[width=7.5cm]{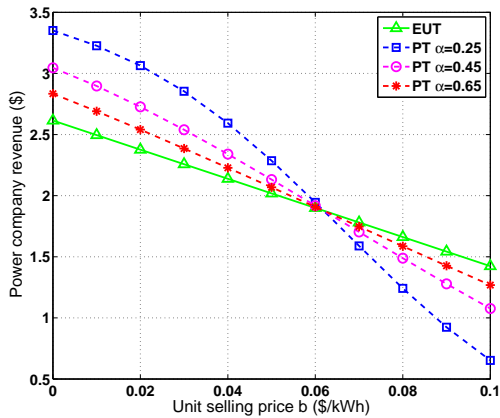}
  \vspace{-0.4cm}
   \caption{\label{fig2} Total revenue for the power company under EUT and PT as the individual customer selling price $b$ varies.}
  \end{center}\vspace{-1cm}
\end{figure}

Fig.~\ref{fig2} evaluates the total revenues of the power company in (\ref{eq:companymoney}) as the customers' unit selling price $b$ increases, for both PT and EUT. %\emph{We assume that the revenues of the power company come only from the payments collected when the customers act as a load} (this does not include the prices from the regulation penalty which can be viewed as either a control parameter or as government taxes/penalties, not collected by the company). 
Fig.~\ref{fig2} clearly shows that, as the unit selling price of the customers increases, the total revenue of the power company will decrease, as the customers start to sell more and buy less. Further, we can clearly see how the deviations from the EUT behavior, as predicted by PT can have a major impact on the market. First, as the customers' unit selling price is below about $\$0.06$ per kWh, under PT, the total revenue collected would be much higher than that expected under EUT. In contrast, if the customers are allowed to set prices that are higher than $\$0.06$ per kWh (and basically higher than the minimum unit price of the company's LMP model), PT predicts that the total revenue will be much smaller than in the EUT case. In this case, it is more beneficial for the power company to regulate the customers' unit selling price to be below $\$0.06$ per kWh (which is comparable to the minimum LMP price of $\$0.05$ per kWh). Fig.~\ref{fig2} demonstrates the importance of incorporating the customers' behavior into the analysis of the power market. In particular, if the power company utilizes EUT to regulate the customers' selling price, in practice, this may incur losses in revenues (relative to EUT) if realistic user behavior models are not accounted for. Finally, we note that the ``crossing point'' between PT and EUT in Fig.~\ref{fig2} depends largely on $\beta$. As $\beta$ becomes higher, a higher unit selling price would be required for the customers to more aggressively sell energy.

%Fig.~\ref{fig4} presents the total average demand from participating players under the impact of unit price $c$. Here, the demand of players is the amount energy they charged. Such amount includes the energy they used and the energy they stored in battery. In this figure, we can see that, the total average of EUT demand is a linear function of unit price due to the product result of objective probability and charging amount. However, the PT demand relies on the subjective observation and each player might determine their actions through considering natural perceptions. This difference explains the variation demand in practice and then, such result can provide the power company with a possible fluctuation when we need to predict local demand. Also, we similarly observe that there has a crossing point of EUT and PT. When the unit price is low/high, the players want to discharge/charge in their mind. This correspondingly explains the phenomenon in practice and depict personal perception about charging/discharging. As a power company, we can have a novel approach to dynamically control local demand: in peak-demand hours, for example, the power company increases the price less than before and reaches a same expected demand. 

\begin{figure}[!t]
 \begin{center}
  \vspace{-0.3cm}
   \includegraphics[width=7.5cm]{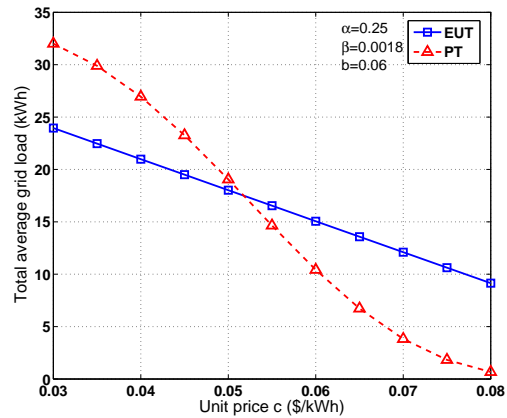}
  \vspace{-0.4cm}
   \caption{\label{fig3} The total average grid load for participating customers under EUT and PT as the power company's unit price $c$ varies.}
  \end{center}\vspace{-0.4cm}
\end{figure}

In Fig.~\ref{fig3}, we show that the expected load on the grid significantly differs between PT and EUT. For PT, when the unit price for buying energy is small, the customers are less interested (compared to EUT) in selling energy now. However, as the unit price crosses a threshold, the customers will sell more aggressively and, thus, the overall load on the grid will be smaller than expected. Fig.~\ref{fig3} can provide important guidelines for demand-side management in the smart grid. For example, assume the power company wants to increase its price to drive customers to sell more and reduce their average load to about $10$ kWh while keeping the generation regulation within limits. Based on EUT, the company would have to increase the minimum LMP price to roughly $\$0.077$ per kWh. In reality, because users behave subjectively when faced with risk, the company does not need to introduce such a high price increase. In contrast, it can increase it to about $\$0.06$ per kWh and obtain the desired load reduction. On the other hand, if the company wants to reduce its price to sustain up to $23$ kWh of load (from the two customers in question), based on EUT, it would have to offer a relatively low price of $\$0.035$ per kWh. In contrast, based on PT, a price of about $\$0.047$ per kWh can achieve the same impact yet yield more profits. Clearly, ignoring the fact that users' behavior can deviate from the rational EUT path can yield undesirable loads on the grid which further motivates the need for PT analysis.

\begin{figure}[!t]
 \begin{center}
  \vspace{-0.2cm}
   \includegraphics[width=7.5cm]{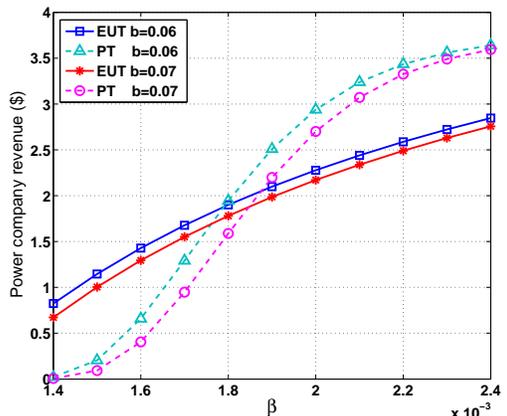}
  \vspace{-0.4cm}
   \caption{\label{fig4} Total revenue for the power company under EUT and PT as the regulation parameter $\beta$ varies.}
  \end{center}\vspace{-1cm}
\end{figure}

In Fig.~\ref{fig4}, we show how the power company revenues under EUT and PT vary as the regulation parameter $\beta$ increases. In particular, we vary $\beta$ from $0.0014$ to $0.0024$ while satisfying the existence of a proper mixed Nash equilibrium. First, the solid lines show that the revenue under EUT is concave. This is due to the fact that the objective probability of charging is computed from a nonlinear utility that integrates power regulation. As the parameter $\beta$ increases, both players want to store/charge more, since discharging increases the penalty of power regulation. %In other words, the increasing $\beta$ decreases the utility of discharging, in which the power company could use such approaches to control the generation level. 
Also, we can see that, after the crossing point (i.e $\beta=0.0018$ when $b=0.06$), the power company would obtain a high revenue from the PT actions of players. This is because players are more likely to charge (act more conservatively) at a high $\beta$ comparable to their objective action. Thus, the power company must choose an optimal $\beta$ while balancing the tradeoff between its own revenues and effective customer participation via discharging. \vspace{-0.15cm}
%Here, we did assume a weight such that the subjective observation might or might not greater than the objective probability in (\ref{eq:weight}). Last but not least, the shape of power company revenue under PT is similar to that Fig~\ref{fig:probaweight}, while Fig.~\ref{fig2} has an inversed shape. This is because, the power company revenue is the players' payment and the players should positively consider the regulation parameter $\beta$ so as to prevent their discharging action from a setting penalty.

\section{Conclusions}\label{sec:conc}\vspace{-0.15cm}

In this paper, we have introduced a novel approach for
studying the problem of customer-owned energy storage integration in the smart grid. We have developed a novel game-theoretic approach, based on prospect theory, using which each player subjectively observes and determines its actions so as to optimize a utility function that captures the benefit from selling energy as well as the associated regulation penalty. For the two-player scenario, we have shown the existence of an equilibrium for both EUT and PT. Simulation results have shown that prospect theory enables the power company to better decide on its pricing parameters, given realistic behavior of the users which deviate considerably from conventional EUT behavior. This paper only scratches the surface of prospect theory, which is expected to become a key technique in the design and analysis of a user-centric smart grid.
%The complex interactions of individual strategies in smart grid to charge/discharge energy. 

%\def\baselinestretch{0.8}
\bibliographystyle{IEEETran}
\bibliography{references}

\end{document}